\documentclass[12pt]{article}

\newcommand{\bea}{\begin{eqnarray}}
\newcommand{\eea}{\end{eqnarray}}
\newcommand{\p}{\partial}
\newcommand{\be}{\begin{equation}}
\newcommand{\ee}{\end{equation}}
\newcommand{\nn}{\nonumber}
\newcommand{\lb}{\label}
\def\trans{\mbox{\tiny$\bot$}} % Transverse component
\def\longi{\mbox{\tiny$\|$}}   % Longitudinal component
\newcommand{\re}[1]{(\ref{#1})}

\begin{document}

\begin{center}
\textbf{ THE KINETIC EQUATION FOR THE QUARK WIGNER FUNCTION\\[0.5ex] IN
STRONG GLUON FIELD}\\[1cm]
{\large A.V. Prozorkevich, S.A. Smolyansky, and S.V. Ilyin}\\[2ex]
\textit{\small Physics Department,
Saratov State University \\
41026  Saratov, Russian Federation\\[0.5ex]
 smol@sgu.ru}
\end{center}

\vspace{0.5cm}
\begin{abstract}
The Vlasov type quantum kinetic equation for the deconfined quarks in the strong
quasi-classical gluon field is derived in the covariant  single-time
formalism. The equations system for the Wigner function components is
obtained as a result of spinor and color decomposition. The field-free and
vacuum solutions of the kinetic equation are found, and the conservation
laws are derived. The flux-tube configuration of gluon field is discussed in
detail.
\end{abstract}

\section{Introduction}

The creation and evolution of the quark-gluon plasma (QGP)  by
ultra-relativistic heavy ion collisions is a very complex process occurring
on the time scale of $1fm/c$ and at energy density above of $1Gev/fm^3$.
Parton gas probably comes to equilibrium state in the process of
hadronization. Details of this transition are of great importance for
interpretation of experimental data. If the QGP is really formed in
experiments at the RHIC, its thermalization time must be so small, that it
presents a serious problem for theoretical explanation \cite{heinz1}.
Several theoretical tools are applied to describe the various physical phenomena
accompanying the collision. The kinetic equation is the basic means for
investigation of the nonequilibrium evolution \cite{wong}-\cite{mark}. The
flux tube model \cite{gatoff} and the Schwinger mechanism of pair production
are often used for the research of an early stage of the QGP formation.
Vacuum pair creation is on essentially non-perturbative effect, which
description requires the exact solution of the field equations. The resulting
source term in the kinetic equation (KE) describes the production rate and
momentum spectra of the created particles \cite{smol1}.

The QED example shows that the evolution of a plasma created from vacuum is
rather dependent on the structure of the source term
\cite{bast}-\cite{smol3}. The semi-phenomenological source term based on the
classical Schwinger formula (e.g., \cite{klug1,klug2}) can result in
significant inaccuracy for the fast varying electrical field. In particular,
such source can not reproduce correctly the relation between the production
rates of two components with different masses and statistics
\cite{vin1,skok}. The main reason is that correct source term has a
non-markovian time dependence \cite{smol1,bast}, as against the Schwinger
formula. The most interesting effects arising from these feature are the
suppression of boson creation with zero kinetic momentum and the suppression
of statistical factor influence. The Schwinger-like source contains the
statistical factor as multiplier and produce the suppression of fermion
creation (Pauli blocking) and enhancement of boson creation. The correct
source term \cite{bast} contains the statistical factor as integrand on time
(non-markovian property) that facilitates the fermion creation  and reduces
the boson creation by large particle density. The joint action of these
factors may cause the effect of "statistic inversion" at the short times
scale when the fermions production rate is more greater than bosons one.
This effect can be so strong that heavy fermions are created more actively
than light bosons.

The other noteworthy feature of a proper source term is that a
momentum distribution of the produced fermion pairs is close  to the
quasi-equilibrium one  both in the transverse and in longitudinal
directions. The influence of such source term can facilitate more rapid
formation of a quasi-equilibrium state of QGP along with the small mean free
path compared to the  Compton length. The correct source term allows to
separate the vacuum polarization effects which can exceed the contribution
of real particles in the thermodynamical variables \cite{vin2}.

The kinetic equation with account of vacuum creation effect can be obtained
in the Wigner function approach, which is widely used for the description of
relativistic quantum systems during the past few decades. It is more
convenient to use the single-time Wigner function variety \cite{bgr,oh} to
solve the initial value problem for the collisions at RHIC. A covariant
version of this approach has been recently developed in \cite{hmr} for the
QED case within the framework of the proper time method on space-like
hyperplanes. We use this approach in the present work for the derivation of
the kinetic equation for the covariant single-time Wigner function (WF) in
the quark sector of QCD at presence of a strong quasi-classical background
gluon field.

The paper is organized as follows. We introduce the equations of motion for
quarks and quasi-classical gluon fields in covariant proper time approach in
Sect. 2. The single-time Wigner function is introduced here as the basic
element of kinetic theory. The physical observables are bilinear
compositions of quark field operators, and can be calculated via WF. The
kinetic equation for the Wigner function is derived in a matrix form in the
spinor and color spaces in Sect. \ref{sec_ke}. The spinor decomposition of this KE is
also fulfilled for the more convenient analysis. This representation of KE
is useful for construction of conservation laws, Sect. \ref{conserv}. The vacuum and
field-free solutions are obtained in Sect.~\ref{secvac}. The vacuum WF play a role of
initial value for a solution of the Cauchy problem. The special sample of
the kinetic equation is investigated in Sect. \ref{homogen}, which is inspired by the
flux-tube \cite{gatoff} model of vacuum quark production under conditions of
ultra-relativistic heavy ion collisions. Comparison with QED case is made in
Sect. \ref{qed}. Finally, Sect. \ref{summ} summarizes some results of the work. We use the
system of units with $\hbar =c=1$ and the signature of metric tensor
$(1,-1,-1,-1)$.

\section{Basic Equations}

We start with the QCD  Lagrange density (for one quark flavor, $g>0$)
\begin{equation}\lb{qcd}
\mathcal{L}_{QCD}=i\bar\psi \gamma^\mu (\partial_\mu
-ig A_\mu )\psi-m\bar \psi\psi-
\frac{1}{2}F_{\mu\nu}F^{\mu\nu},
\end{equation}
with the quasi-classical gluon field
\begin{equation}\label{field}
A_\mu = A_\mu^a \, t^a,\qquad
F_{\mu\nu}=\partial_\mu A_\nu -\partial_\nu
A_\mu -ig\,[A_\mu ,A_\nu ],\qquad a=1,2...N,
\end{equation}
where $t^a=\lambda^a/2$ are generators of SU(N) gauge group in fundamental
representation. In particular, $\lambda^a$ are the Gell-Mann matrices for SU(3)
and the Pauli ones for the SU(2) group.

The covariant proper time derivative $\partial_\tau$
is defined by means of the space-time parametrization by a family of
space-like hyperplanes $\sigma(n,\tau)$
\be\label{sigma}
\sigma(n,\tau)\bot\ n^\mu, \qquad n^\mu x_\mu=\tau,
\qquad n^2=1.
\ee
Hence there is a covariant decomposition
\bea
x^\mu=\tau n^\mu +x^\mu_{\trans},\qquad
\partial_\mu = n_\mu\partial_\tau +\partial_\mu^{\trans},\nn\\
\p_\tau = n^\mu \p_\mu,\qquad
\partial_\mu^{\trans}=\Delta_{\mu\nu}\p^\nu,\qquad
\Delta^{\mu\nu}=g^{\mu\nu}-n^\mu n^\nu,
\eea
where $n^\mu$ is a  unit time-like vector,  $\Delta^{\mu\nu}$ denotes the
transverse projector, and $g^{\mu\nu}$ is the metric tensor.
Analogous decomposition is produced for any vector $f^\mu$
\be\label{decomp}
f^\mu = f^{\longi} n^\mu +f^\mu_{\trans},\qquad f^{\longi}= f^\mu n_\mu,
\qquad f^\mu_{\trans}=\Delta^\mu_\alpha f^\alpha,
\ee
and for any anti-symmetric tensor $c^{\mu\nu}$
\begin{equation}\label{cmu}
c^{\mu\nu}= c^\nu n^\mu -c^\mu n^\nu
+c^{\mu\nu}_{\trans}, \qquad c^\mu =n_\nu c^{\nu\mu},
\qquad c^{\mu\nu}_{\trans}=\Delta^\mu_\alpha\Delta^\nu_\beta
c^{\alpha\beta},
\end{equation}
so $c_\mu^{\trans}=c_\mu$.
In these terms, the basic equations of motion for quark field operators
are
\bea\label{dirac}
\partial_\tau \psi &=& -\gamma^{\longi}\gamma^\mu_{\trans}
(\partial_\mu^{\trans} -ig A_\mu^{\trans})\psi -im\gamma^{\longi}\psi
+ig A^{\longi}\psi, \nn\\
\partial_\tau \bar\psi &=& -\bar\psi
(\stackrel{\leftarrow}{\partial_\mu^{\trans}}
+ig A_\mu^{\trans})\gamma^\mu_{\trans}\gamma^{\longi}
+im\bar\psi\gamma^{\longi}
-ig \bar\psi A^{\longi}.
\end{eqnarray}
The  mean gluon fields obey the equations
\begin{eqnarray}
\p_\tau E_\nu&=&-\p^\mu_{\trans}F_{\mu\nu}^{\trans}+
ig\left([A^\mu_{\trans},F_{\mu\nu}^{\trans}]-[A^{\longi},E_\nu]\right)
-gJ^{\trans}_\nu,\nn \\
\p^\mu_{\trans}E_\mu &=&-ig[A^\mu_{\trans},E_\mu]+gJ^{\longi},
\end{eqnarray}
where $E_\nu=n^\mu F_{\mu\nu}$ and $F_{\mu\nu}^{\trans}$
represent the color electrical and
color magnetic fields, respectively, and $J_\nu$ is a color current
\begin{equation}\label{tok}
  J_\nu=J^a_\nu t^a,\qquad
 J^a_\nu = \,<\bar\psi\gamma_\nu t^a\psi>_{\sigma},
\end{equation}
the symbol $<...>_{\sigma}$  designates the average with
statistical operator of the system on a hyperplane
in the Heisenberg picture.

The basic element of statistical description is the covariant
Wigner function (WF)
on the space-like hyperplane  $\sigma(\tau)$ \cite{bgr,hmr}:
\begin{eqnarray}\label{wf}
W^{ab}_{ik}(x_{\trans},p_{\trans};\tau)&=&
 \int d^4 y\, {\rm e}^{ip\cdot y}\,\delta(y\cdot n)\times \nn\\
&\times & U^{aa'}(x,x+y/2)\,
\rho^{a'b'}_{ik}(x,y;\tau)\,
U^{b'b}(x-y/2,x),
\end{eqnarray}
here the upper and bottom indices are the color and spinor ones, respectively,
$\rho$ is the one-particle density matrix of quarks
\be\label{densm}
\rho^{ab}_{ik}(x,y;\tau)=-\frac12
<\left[\psi^a_i(x+y/2),
\bar\psi^b_k(x-y/2)\right]>_\sigma,
\ee
and $U$ is the unitary link operator
\be
U(x_1,x_2)=\exp{\left(ig\int\limits_{x_1}^{x_2} dz^\mu A_\mu(z)\right)},
\ee
provides the gauge invariance of the WF \cite{oh}.
The integral is taking along straight line between points $x_1$
and $x_2$ connected with space-like interval and
  \begin{eqnarray}\label{link}
   U^+(x_1,x_2)=U^{-1}(x_1,x_2)=U(x_2,x_1), \nn\\
    U(x_1,x_2)\,U(x_2,x_1)=U(x_1,x_1)=\hat 1.
  \end{eqnarray}

The dynamical variables can be expressed in terms of the Wigner function.
For example, the color and electromagnetic  current densities are
\bea\label{tokc}
  J^a_\nu (x) &=& \mathrm{Tr} < \gamma_\nu t^a W>, \\
\label{tokb}
j_\nu (x) &=& \mathrm{Tr} < \gamma_\nu W>,
\eea
where  trace $\mathrm{Tr}$ is carried out on spinor and color indices.
The brackets $<...>$ denote the covariant
momentum average on the hyperplane $p\cdot n=0$ \cite{hmr}, e.g.
\begin{equation}\label{mom}
<W>=\int\frac{d^4 p}{(2\pi)^3}\,\delta(p\cdot n)\,
W(x_{\trans},p_{\trans};\tau =x\cdot n).
\end{equation}
The integral in \re{mom} provides the locality of
corresponding observables, due to that function
\begin{equation}\label{delta}
\delta_\tau (x)= \int\frac{d^4 p}{(2\pi)^3}\,
e^{-ipx}\, \delta(p\cdot n)
\end{equation}
plays the role of the three-dimensional delta function on a hyperplane
$\sigma(n,\tau)$.

The color WF have a rather complex matrix structure,
therefore it is convenient to use the corresponding decompositions in the
spinor and color space. The spinor decomposition in a complete basis of the
Clifford algebra is
\be\lb{spinor}
W=aI+b_\mu\gamma^\mu+c_{\mu\nu}\sigma^{\mu\nu}+
d_\mu\gamma^\mu\gamma^5+ie\gamma^5,
\ee
where the coefficient functions $a$ (scalar), $b_\mu$ (vector),
$c_{\mu\nu}$ (antisymmetric tensor), $d$ (axial vector), $e$ (pseudo-scalar)
are the hermitian color matrices
\begin{eqnarray}\label{w_ik}
a=\frac{1}{4}\,\mathrm{tr}\,W,\quad
b^\mu=\frac{1}{4}\,\mathrm{tr}\,(\gamma^\mu W),\quad
c^{\mu\nu}=\frac{1}{8}\,\mathrm{tr} (\sigma^{\mu\nu}W), \nn\\
d^\mu=-\frac{1}{4}\,\mathrm{tr}\,(\gamma^\mu\gamma^5 W),\qquad
e=-\frac{i}{4}\,\mathrm{tr}\, (\gamma^5 W),
\end{eqnarray}
the symbol "$\mathrm{tr}$" denotes the trace on spinor indices only,
and $\gamma^5=-i\gamma^0\gamma^2\gamma^2\gamma^3$.

The algebraic structure of the WF in a color space of $N\times N$ matrices
is represented by the observable color singlet $W^a$ and the unobservable
color multiplet $W^a$,
\begin{eqnarray}\label{col}
W=W^s\cdot \hat 1 + W^a\cdot t^a, \qquad a=1,2\ldots (N^2-1),\\
W^s= (1/N)\, \mathrm{tr_c}\, W,\qquad
W^a= 2\, \mathrm{tr_c} \,(t^a W),\nn
\end{eqnarray}
where $\hat 1$ is unit matrix and $\mathrm{tr_c}$
is the trace on color indices.

\section{Kinetic Equation}\label{sec_ke}
We calculate the proper time derivative of Eq.\re{wf}
for a derivation of the KE for the Wigner function
  \begin{eqnarray}\label{der}
\p_\tau W^{ab}_{ik} =
 \int d^4 y\, {\rm e}^{ip\cdot y}\,\delta(y\cdot n)\left\{
U^{aa'}(x,x+y/2)\,U^{b'b}(x-y/2,x)
\p_\tau \rho^{a'b'}_{ik}+\right.\nn\\ \left.
+\rho^{a'b'}_{ik}
\p_\tau\left[U^{aa'}(x,x+y/2)\,U^{b'b}(x-y/2,x)
\right]\right\},
\end{eqnarray}
and substitute the time derivative of the field operators from
the field equations \re{dirac}. The emerging terms with perpendicular
derivative $\p_{\trans}$ are transformed via the integration by parts.
The derivative rules for link operator are follow from known formula
\cite{egv,oh}
\begin{eqnarray}\label{du}
\delta U(x_1,x_2)= ig\delta x_1^\mu A_\mu(x_1)U(x_1,x_2)
-ig U(x_1,x_2)A_\mu(x_2)\delta x_2^\mu-\nn\\
-ig\int\limits_0^1 ds U(x_1,z(s))F_{\mu\nu}(z(s))U(z(s),x_2)
(x_1-x_2)^\mu [\delta x_2 +s(\delta x_1 -\delta x_2)]^\nu,
\end{eqnarray}
where $z(s)=x_2+s(x_1-x_2)$, e.g.
\begin{eqnarray}\label{linkder}
\p_\mu(x) U(x,x+y/2)= - ig \,\biggr[\,A^{[x]}_\mu(x,x+y/2)
-A_\mu(x)+ \nn\\
+\int\limits_0^{1/2} ds\, F^{[x]}_{\mu\nu} (x+sy)\, y^\nu\
\biggr]\,U(x,x+y/2),
\end{eqnarray}
where the notation for the "Schwinger string" \cite{oh}
\begin{equation}\label{string}
A^{[x]}_\mu (z)=U(x,z)\,A_\mu (z)\,U(z,x)
\end{equation}
is introduced.
Then we express the variable $y$ as the momentum derivative
\begin{equation}\label{derp}
y^\mu\exp{(ip\cdot y)}\delta(y\cdot n)=
\delta(y\cdot n)\p^\mu_{\trans}(p)\,
\exp{(ip_{\trans}\cdot y_{\trans})},
\end{equation}
and we obtain after some algebra the exact equation of motion for the WF
\begin{eqnarray}\lb{ke}
\p_\tau W +\frac12\p^{\trans}_\mu(x)[S^\mu,W]+
im[\gamma^\parallel,W]-ip^{\trans}_\mu \{S^\mu,W\}+\nn\\
+g\p^\mu_{\trans}(p)\!\int\limits_0^{1/2}\!ds
\left\{E^{[x]}_\mu(\tau,x_{\trans}\!-is\p_{\trans}(p))W+
W E^{[x]}_\mu (\tau,x_{\trans}\!+is\p_{\trans}(p))\right. +\nn\\
+\frac{1}{2}F^{[x]}_{\mu\nu}\left(\tau,x_{\trans}-is\p_{\trans}(p)\right)
\biggl([W,S^\nu]-2s\{W,S^\nu\}\biggr)+\nn\\
\left.+
\frac{1}{2}\biggl([W,S^\nu]+2s\{W,S^\nu\}\biggr)
F^{[x]}_{\mu\nu}\left(\tau,x_{\trans}+is\p_{\trans}(p)\right)
\right\}=\nn\\
= ig[A_\parallel(x),W]+
\frac{ig}{2}\left[S^\mu , [A_\mu^{\trans},W]\right],
\eea
where
\be
S^\mu=\gamma^{\longi}\gamma^\mu_{\trans},\qquad
\gamma^{\longi}=n^\mu \gamma_\mu,\qquad
\gamma^\mu_{\trans}=\Delta^\mu_\nu\gamma^\nu.
\ee
This equation is resulted in a local form by means of
the  gradient expansion of a gluon field. We obtain in the
first oder of this procedure the kinetic equation of the Vlasov  type,
which is correct for a rather slowly changing gluon field
\bea\lb{lke}
\p_\tau W+\frac12
\p^{\trans}_\mu(x)[S^\mu,W]+\nn \\
+\frac{g}{8}\p^\mu_{\trans} (p)\left( 4\{W,E_\mu\}+
2\left\{F_{\mu\nu},[W,S^\nu]\right\}-
\left[F_{\mu\nu},\{W,S^\nu\}\right] \right)=\nn\\
=ip^{\trans}_\mu \{S^\mu,W\}-im[\gamma^\parallel,W]
+ig\left([A_\parallel(x),W]+\left[A_\mu^{\trans}\,,
 [S^\mu ,W]\right]\right)\,.
\end{eqnarray}
This equation despite of rather compact form is a very complicated matrix
one in the direct production of $4\times 4$ spinor and $N\times N$ color
space. It is convenient to expand the WF in a some basis of this space for
the separation of a different physical contributions.

We perform at first the spinor decomposition \re{spinor} in the KE. We
obtain the equations set calculating the traces of Eq.\re{lke} with the
basic matrices of Clifford algebra
\begin{eqnarray}\lb{spura} %with I
\p_\tau a+\frac{g}{2}\p^\alpha_{\trans} (p)\left(\{a,E_\alpha\}+
i[F^{\trans}_{\alpha\beta},c^\beta_{\trans}]\right)=
4p^\alpha_{\trans} c^{\trans}_\alpha+ig[A^\parallel,a],\\ \nn\\
% with gamma
\p_\tau b^\mu+ n^\mu \p_\nu^{\trans} b^\nu_{\trans}-\p^\mu_{\trans} b^\parallel+\nn\\
+\frac{g}{2}\p^\alpha_{\trans} (p)\left(\{b^\mu,E_\alpha\}+
g^{\mu\beta}\{b^\parallel,F_{\alpha \beta}^{\trans}\}
-n^\mu\{b_{\trans}^\beta,F_{\alpha\beta}^{\trans}\}+
\frac{i}{2}[F_{\alpha\beta}^{\trans},d_\delta]
\varepsilon^{\,\beta\delta\mu}\right)=\nn\\
=2p_\alpha^{\trans} d_\beta^{\trans} \varepsilon^{\,\alpha\beta\mu}+4mc_{\trans}^\mu
+ig\left([A^\parallel,b^\mu]+n^\mu[A_{\trans}^\alpha,b_\alpha^{\trans}]
-[A_{\trans}^\mu,b^\parallel]\right),\lb{bmu}\\ \nn\\
%with sigma
\p_\tau c^{\mu\nu}+\p_\alpha^{\trans} (n^\nu c^{\mu\alpha}_{\trans}-
n^\mu c^{\nu\alpha}_{\trans})+\p_{\trans}^\nu c^\mu -\p_{\trans}^\mu c^\nu+\nn\\
+\frac{g}{2}\p^\alpha_{\trans}(p)\left(\{E_\alpha,c^{\mu\nu}\}
+\{F_{\alpha\beta}^{\trans},
(n^\mu c^{\nu\beta }_{\trans}-n^\nu c^{\mu\beta}_{\trans}+g^{\mu\beta} c^\nu
-g^{\nu\beta}c^\mu)\}\right.+\nn\\+\left.\frac{i}{4}\left[F_{\alpha\beta}^{\trans},
a(n^\mu g^{\nu\beta}-n^\nu g^{\mu\beta})+
e\varepsilon^{\beta\mu\nu}
\right]\right)=\nn\\ =
a(n^\mu p^\nu_{\trans}-n^\nu p^\mu_{\trans})
+ep^{\trans}_\alpha\varepsilon^{\,\alpha\mu\nu}
+m(n^\nu b^\mu-n^\mu b^\nu)+\nn\\ \nn\\
+ig[A^\parallel,c^{\mu\nu}]+ig\left[A_\alpha^{\trans},
(n^\nu c^{\mu\alpha}_{\trans}-n^\mu c^{\nu\alpha}_{\trans}+g^{\nu\alpha} c^\mu
-g^{\mu\alpha}c^\nu)\right],\\ \nn\\
% with gamma^5 gamma
\p_\tau d^\mu-\p_{\trans}^\mu d^{\,\parallel}+n^\mu\p_{\trans}^\alpha
 d^{\trans}_\alpha+\nn\\+
\frac{g}{2}\p^\alpha_{\trans}(p)
\left(\{d^\mu,E_\alpha\}+g^{\mu\beta}\{F_{\alpha \beta}^{\trans},d^{\,\parallel}\}
-n^\mu\{F_{\alpha\beta}^{\trans},d^\beta_{\trans}\}
+\frac{i}{2}[F_{\alpha\beta}^{\trans},b_\delta]
\varepsilon^{\,\beta\delta\mu}\right)=\nn\\=
2p_\alpha^{\trans} b_\beta\varepsilon^{\,\alpha\beta\mu}+
2m n^\mu e+ig\left([A^\parallel,d^\mu]+
n^\mu[A^\alpha_{\trans},d_\alpha^{\trans}] -
[A^\mu_{\trans},d^{\,\parallel}]\right),\\ \nn\\\lb{spure}
% with gamma^5
\p_\tau e+\frac{g}{2}\p^\alpha_{\trans} (p)\left( \{e,E_\alpha\}-\frac{i}{2}
[F_{\alpha\beta},c_{\delta\rho}]
\varepsilon^{\,\beta\delta\rho}\right)=\nn\\ = -2p_\alpha
c_{\beta\delta}\varepsilon^{\,\alpha\beta\delta}
-2md^{\,\parallel}+ig[A^\parallel,e],
\end{eqnarray}
where $\varepsilon^{\,\alpha\beta\gamma}$ denote the convolution of the
normal $n^\mu$ with the totally anti-symmetric unit tensor
$\varepsilon^{\,\alpha\beta\gamma\delta}$, i.e.
\begin{equation}
\varepsilon^{\,\alpha\beta\gamma}=n_\rho
\varepsilon^{\,\rho\alpha\beta\gamma},\qquad \varepsilon^{\,0123}=+1.
\end{equation}
It is more convenient  for the analysis of concrete field configurations to
rewrite that system concerning a projections on time-like and space-like
directions
\begin{eqnarray}\label{proj_a}
\p_\tau a+\frac{g}{2}\p^\alpha_{\trans} (p)\left(\{a,E_\alpha\}+
i[F_{\alpha\beta}^{\trans},c^\beta_{\trans}]\right)= 4p^\alpha_{\trans}
c^{\trans}_\alpha+ig[A^\parallel,a],\lb{scalar}\\
% b_parallel
\lb{vec_par}
\p_\tau b^\parallel+\p^\alpha_{\trans} (x) b_\alpha^{\trans}
+\frac{g}{2}\p^\alpha_{\trans} (p)\left(\{b^\parallel,E_\alpha\}
-\{b^\beta_{\trans},F_{\alpha\beta}^{\trans}\}\right)=ig[A^\alpha,b_\alpha],
\eea \bea
% b_perp
\lb{vec_perp}
\p_\tau b^\mu_{\trans}
+\frac{g}{2}\p^\alpha_{\trans} (p)\left(\{b^\mu_{\trans},E_\alpha\}+
\{b^\parallel,F_{\alpha }^{{\trans} \mu} \}+
\frac{i}{2}[F_{\alpha\beta}^{\trans},d_\delta^{\trans}]
\varepsilon^{\,\beta\delta\mu}\right)=\nn\\
= \p_{\trans}^\mu (x) b^\parallel+
2p_\alpha^{\trans} d_\beta^{\trans}\varepsilon^{\,\alpha\beta\mu}+4mc_{\trans}^\mu
+ig\left([A^\parallel,b^\mu_{\trans}]
-[A_{\trans}^\mu,b^\parallel]\right),\eea \bea
% tensor_parallel
\p_\tau c_{\trans}^\mu + \p_\alpha^{\trans} (x) c^{\alpha\mu}_{\trans} +
\frac{g}{2}\p^\alpha_{\trans} (p)\left(\{E_\alpha,c_{\trans}^\mu\}-
\{F^{\trans}_{\alpha\beta},
c_{\trans}^{\beta\mu}\}+\frac{i}{4}
[F^{{\trans}\mu}_{\alpha},a]\right)=\nn\\
=p_{\trans}^\mu a- m b_{\trans}^\mu+ig\left(
[A^\parallel,c_{\trans}^\mu]+[A_\alpha^{\trans},
c_{\trans}^{\alpha\mu}]\right),\eea
%tensor_perp
\bea
\p_\tau c^{\mu\nu}_{\trans}+\p^\nu_{\trans} (x) c_{\trans}^\mu-
\p^\mu_{\trans} (x) c_{\trans}^\nu +\nn\\
+\frac{g}{2}\p^\alpha_{\trans}(p)\left(\{E_\alpha,c^{\mu\nu}_{\trans}\}+
\{F_{\alpha}^{{\trans}\mu},c_{\trans}^\nu\}
-\{F_{\alpha}^{{\trans}\nu},c_{\trans}^\mu\}+
\frac{i}{4}[F_{\alpha\beta}^{\trans},e]\,
\varepsilon^{\,\beta \mu\nu}\right) =\nn\\
=e p_\alpha^{\trans} \,\varepsilon^{\alpha\mu\nu}
+ig\left([A^\parallel,c^{\mu\nu}_{\trans}]+[A^\nu_{\trans},c_{\trans}^\mu]
-[A^\mu_{\trans},c_{\trans}^\nu]\right),\eea \bea
% axial parallel
\p_\tau d^\parallel +\p_{\trans}^\alpha (x) d^{\trans}_\alpha +
\frac{g}{2}\p^\alpha_{\trans}(p)
\left(\{d^\parallel,E_\alpha\}-\{F_{\alpha\beta}^{\trans},d^\beta_{\trans}\}
\right)=\nn \\ =2me+ig\left([A^\parallel,d^\parallel]+
[A^\alpha_{\trans},d_\alpha^{\trans}]\right),\eea \bea
% axial perp
\p_\tau d^{\,\mu}_{\trans}+
\frac{g}{2}\p^\alpha_{\trans}(p)
\left(\{d^{\,\mu}_{\trans},E_\alpha\}+
\{F_{\alpha}^{{\trans}\mu},d^{\,\parallel}\}
+\frac{i}{2}[F_{\alpha\beta}^{\trans},b^{\trans}_\delta]
\varepsilon^{\,\beta\delta\mu}\right)=\nn\\=
\p_{\trans}^\mu (x) d^{\,\parallel}+
2p_\alpha^{\trans} b_\beta^{\trans}\varepsilon^{\,\alpha\beta\mu}
+ig\left([A^\parallel,d^{\,\mu}_{\trans}]
-[A^\mu_{\trans},d^{\,\parallel}]\right),\eea
% pseudo-scalar
\bea \p_\tau e+\frac{g}{2}\p^\alpha_{\trans} (p)\left(
\{e,E_\alpha\}-\frac{i}{2}
[F_{\alpha\beta}^{\trans},c_{\delta\rho}^{\trans}]
\varepsilon^{\,\beta\delta\rho}\right)=\nn\\ = -2p_\alpha^{\trans}
c_{\beta\delta}^{\trans}\varepsilon^{\,\alpha\beta\delta}
-2md^{\,\parallel}+ig[A^\parallel,e].\lb{proj_e}
\end{eqnarray}
The non-abelian character of gluon field is displayed in particular in the
presence of the vector potential in right-hand side of these KEs that is
necessary for the gauge invariance of the theory.

It is the resulting set of KE for the description of deconfined quarks in a
strong quasi-classical gluon field.

\section{Field-Free Limit and Vacuum Solution}\label{secvac}

The equations system \re{proj_a}-\re{proj_e} can be solved exactly in the
field-free limit $A_\mu=0$. We obtain assuming that all derivatives vanish:
\begin{eqnarray}\label{free}
 c_\mu=0,\qquad c_{\mu\nu}^{{\trans}}=0,\qquad d_\mu=0,
 \qquad e=0 ,\nn\\
 p_{{\trans}}^\mu a(p)-m b_{{\trans}}^\mu (p) = 0, \nn\\
p_\alpha^{\trans} b_\beta^{{\trans}}\varepsilon^{\,\alpha\beta\mu}=0.
\end{eqnarray}
The general solution of \re{free} is
\begin{equation}\label{freesol}
b_{{\trans}}^\mu (p)=p_{{\trans}}^\mu \frac{a(p)}{m}
\end{equation}
with an arbitrary momentum dependence of $a(p)$. The equilibrium WF
\cite{hakim} or the vacuum solution follow from the Eq.\re{freesol} as a
particular cases. We perform the direct calculation of the function \re{wf}
to select the vacuum case in the solutions class \re{freesol}. The fermion
operators on the hyperplane obey free anticommutation relations
\cite{hmr,ann}
\begin{equation}\label{anticom}
\{\psi^a(x_2),\bar\psi^b(x_1)\}=\delta^{ab}\,
\gamma^{\longi}\delta_\tau(x_1-x_2).
\end{equation}
Then allows to write down the commutator \re{densm} as
\begin{eqnarray}\label{::}
\rho^{ab}(x,y;\tau)= <:\psi^a(x_2)\bar\psi^b(x_1):>_\tau
 - \nn \\
-\frac{1}{2}\,\delta^{ab}\int\frac{d^4k}{(2\pi)^3}e^{ik_{\trans}\cdot
y_{\trans}}\delta(k^2-m^2)(m-k\cdot\gamma)
\left[\theta(k^{\longi})+\theta(-k^{\longi}\right],
\end{eqnarray}
where the symbol $:\ :$ indicates normal ordering. We obtain the vacuum WF
by carrying out the averaging on vacuum state and  substituting this
representation in Eq.\re{wf}
\begin{equation}\label{vac}
  W^{ab}_{vac}=\frac{1}{2}\,\delta^{ab}
  \left(\frac{-m+p_{\trans}\cdot\gamma_{\trans}}
{\omega(p_{\trans})}\right),
\end{equation}
where $\omega(p_{\trans})=\sqrt{m^2-p_{\trans}^2}$.
This function is degenerated  in the color space as a
consequence of the primary supposition about the Lagrange density
structure \re{qcd}.
The WF \re{vac} is the base for a solution of the Cauchy problem for the
equations system \re{proj_a}-\re{proj_e}.
It can be proved that the vacuum solution, as well as the free-field ones,
gives no contribution to the current densities \re{tokc},\re{tokb}.

The kinetic equation has in the field-free case besides the vacuum solution
\re{vac} the another solutions of the type
\begin{equation}\label{quasivac}
W^a(p)=\eta^a\, W^s(p), \qquad a=1,2\ldots N,
\qquad W=\{a, b_{{\trans}}^\mu, 0 \ldots 0\},
\end{equation}
where $\eta^a$ is an arbitrary real numbers.

\section{Conservation Laws}\label{conserv}

We calculate the divergence of electromagnetic current \re{tokb}, using the
spinor and color decompositions \re{spinor} and \re{col}
\begin{equation}
\p^\mu (x)j_\mu(x) = 4N \p^\mu (x)< b_\mu^s(x) > = 4N[\p_\tau\! <
b^s_{\longi}> +\,\p^\mu_{\trans}\!<b^{\trans s}_\mu>].
\end{equation}
Performing the momentum averaging procedure \re{mom} and taking the trace in
Eq.\re{vec_par}, we obtain the electromagnetic current conservation law
\begin{equation}
\p^\mu (x)j_\mu(x) = 0.
\end{equation}
We multiply the equation \re{vec_par} on matrix $t^a$ and repeat the same
procedure  to derive the equation for the color current
$J^a_\mu=(1/2)<b^a_\mu>$. The result is
\be
\p^\mu (x)J^a_\mu(x) +gf^{abc} A^b_\mu J^{\mu c} = 0.
\ee
The energy density $\varepsilon_q$ of quark matter corresponding
Eq.\re{qcd} is
\begin{equation}\label{en1}
\varepsilon_q \equiv n_\mu n_\nu T_q^{\mu\nu}=
i:\bar\psi \gamma^{\longi}(\p_\tau -igA^{\longi})\psi:,
\end{equation}
where $T_q^{\mu\nu}$ - energy-momentum tensor.
Using  the field equations \re{dirac}, we have
\begin{equation}\label{en2}
\varepsilon_q =:i\bar\psi\gamma^{{\trans}}(\p_{{\trans}}-
igA_{{\trans}})\psi+m\bar\psi\psi:,
\end{equation}
This variable
can be written in terms of the WF  as
\be
\varepsilon_q = \mathrm{Tr} < (m - \gamma^{\trans}p_{\trans})W> +
\varepsilon_{vac}, \ee where $\varepsilon_{vac}=2N<\omega(p_{\trans})>$ is
divergent vacuum contribution. We obtain using the spinor and color
decompositions
\begin{equation}
\varepsilon_{M}=4N[m<a^s>-<p^\mu_{\trans} b^{\trans s}_\mu>]
+\varepsilon_{vac}.
\end{equation}
The linear combination of Eqs.\re{proj_a} and \re{vec_perp} is used to
calculate the right side of this equation. The result is
\be
\p_\tau \varepsilon_{M} + 4\p^\mu_{\trans}(x)
<p_\mu^{\trans} b^s_{\longi}> + g J_\mu^{\trans a} E^{\mu
a}_{\trans}=0.
\ee

\section{Space-Homogeneous Color Field}\label{homogen}

\subsection{Flux-tube field configuration}
We consider here the "instant" frame of reference where $n^\mu=(1,0,0,0)$
and the field configuration  typical for the flux tube model
\begin{equation}\label{electr}
 A_\mu^a=(0,0,0,A^a(t)),\qquad F^a_{03}=-\dot
A_3^a(t)=E_3^a(t)=E^a(t),
\end{equation}
where the dot denotes the derivative with respect to time $\tau=t$ and the
Hamilton gauge was selected. We have $\mathbf{E}=(0,0,E)$,  $\mathbf{H}=0$ in
3-vector representation
\be
F^{\mu\nu}=(-\mathbf{E}, \mathbf{H}),\qquad c^{\mu\nu}=(-\mathbf{c}_1,
\mathbf{c}_2), \quad c^\mu=(0,-\mathbf{c}_1).
 \end{equation}
We limit oneself to simple sample of $SU(2)$ group ($a=1,2,3)$ below, where
\be
\{t^a,t^b\}=\frac{1}{2}\delta^{ab}\hat 1,\qquad
[t^a,t^b]=if^{abc}t^c, \qquad \mathrm{tr_c}(t^a t^b)=\frac{1}{2}
\delta^{ab},
\ee
the totally anti-symmetric
structure constants $f^{abc}$ coincide here with
the totally anti-symmetric unit 3-tensor $e^{ijk}$,
$f^{123}=+1$.
As a result of color decomposition of the WF,
the system \re{proj_a}-\re{proj_e} is reduced to
($\p_p=\p /\p p_3$)
\begin{eqnarray}\label{hom}
\p_t a^s+\frac{g}{4}E^a\p_p a^a &=& 4\mathbf{p} \mathbf{c}_1^s,\nn\\
\p_t a^a+gE^a\p_p a^s&=&4\mathbf{p}\mathbf{c}_1^a,\nn\\
\p_t b_0^s+\frac{g}{4}E^a\p_p b_0^a &=& 0,\nn\\
\p_t b_0^a+g E^a\p_p b_0^s &=& -g f^{abc}\mathbf{A}^b \mathbf{b}^c,\nn\\
\p_t \mathbf{b}^s +\frac{g}{4}E^a\p_p \mathbf{b}^a &=& 2\mathbf{p}\times\mathbf{d}^s+
4m\mathbf{c}_1^s,\nn\\
\p_t \mathbf{b}^a+gE^a\p_p \mathbf{b}^s &=& 2\mathbf{p}\times\mathbf{d}^a
-4m\mathbf{c}_1^a
+g f^{abc}\mathbf{A}^b \mathbf{b}_0^c,\nn\\
\p_t \mathbf{c}_1^s +\frac{g}{4}E^a\p_p \mathbf{c}_1^a &=&
-\mathbf{p} a^s +m \mathbf{b}^s,\nn\\
\p_t\mathbf{c}_1^a+gE^a\p_p \mathbf{c}_1^s &=& -\mathbf{}p a^a + m\mathbf{b}^a
+g f^{abc}(\mathbf{A}^b\times \mathbf{c}_2^c),\nn\\
\p_t \mathbf{c}_2^s+\frac{g}{4}E^a\p_p \mathbf{c}_2^a &=& -\mathbf{p} e^s,\nn\\
\p_t\mathbf{c}_2^a+ gE^a\p_p \mathbf{c}_2^s &=& -\mathbf{p} e^a +
g f^{abc}(\mathbf{A}^b\times \mathbf{c}_1^c),\nn\\
\p_t d_0^s+\frac{g}{4}E^a\p_p d_0^a &=& 2me^s,\nn\\
\p_t d_0^a+ gE^a\p_p \mathbf{d}_0^s &=& 2m e^a +
g f^{abc}\mathbf{A}^b \mathbf{c}_1^c,\nn\\
\p_t \mathbf{d}^s+\frac{g}{4}E^a\p_p \mathbf{d}^a &=& 2\mathbf{p}\times\mathbf{b}^s,\nn\\
\p_t \mathbf{d}^a+ gE^a\p_p \mathbf{d}^s &=& 2\mathbf{p}\times \mathbf{b}^a+
g f^{abc}\mathbf{A}^b \mathbf{d}_0^c,\nn\\
\p_t e^s+\frac{g}{4}E^a\p_p e^a &=& 4\mathbf{p}\mathbf{c}_2^s -2m d_0^s,\nn\\
\p_t e^a+\frac{g}{4}E^a\p_p \mathbf{e}^s &=& 4\mathbf{p}\mathbf{c}_2^a-2md_0^a.
\end{eqnarray}
The important difference of these equations from their QED analogue is the
structure of a force terms: the derivatives on time and on momentum act on
different parts of WF (singlet and multiplet, respectively). This feature
does not allow to reduce the problem to solving of ordinary differential
equations even for the simple field configuration \re{electr}. The situation
becomes complicated even more in case of SU(3) ($a=1,2,\ldots,8)$  where
\begin{equation}\label{su3}
\{t^a,t^b\}=\frac{1}{3}\delta^{ab}\hat 1+d^{abc}t^c,
\end{equation}
and $d^{abc}$ are the totally symmetric structure constants. We shall write
out for an illustration two first the equations \re{hom} only:
\bea\label{hom3} \p_t a^s+\frac{g}{6}E^a\p_p a^a &=& 4\mathbf{p}
\mathbf{c}_1^s,\nn\\ \p_t a^a+gE^a\p_p a^s+g d^{\,abc}E^b \p_p a^c
&=&4\mathbf{p}\mathbf{c}_1^a. \eea

\subsection{Constant Chromo-Magnetic Field}

Now we consider the gluon field configuration corresponding to a space-time
homogeneous field
\be
A^\mu=(0,\mathbf{A}),\qquad F^{\mu\nu}=(0,\mathbf{H}),\qquad
\mathbf{A}=const,\quad \mathbf{H}=const. \ee
In this case the system
\re{scalar}-\re{proj_e} is reduced to two independent groups of the equations
($\mathbf{D}=g\mathbf{H}\times \p_{\mathbf{p}}$)
\bea \{\mathbf{D},\mathbf{b}\} &=& 2ig[\mathbf{A},\mathbf{b}\,],\nn\\
i[\mathbf{D}, \mathbf{c}_2]&=& -8\mathbf{p} \,\mathbf{c}_2 + 4md^{\,0}.\nn\\
\{\mathbf{D}\times,\mathbf{c}_2\}+\frac{i}{4}[\mathbf{D},a]&=&
2\left(\mathbf{p} a -m\mathbf{b} +ig[\mathbf{A}\times,\mathbf{c}_2]\right),\nn\\
\{\mathbf{D},d^0\, \}+\frac{i}{2}[\mathbf{D}\times ,\mathbf{b}\,] &=&
-4\mathbf{p}\times\mathbf{b} -2ig[\mathbf{A},d^{\,0}],\lb{c2}\\
\{\mathbf{D},\mathbf{d}\ \}&=&-4me+2ig[\mathbf{A},\mathbf{d}\,],\nn\\
i[\mathbf{D} ,\mathbf{c}_1]&=& -8\mathbf{p} \,\mathbf{c}_1,\nn\\
-i[\mathbf{D}\times ,\mathbf{d}\,] &=& 8\mathbf{p}\times\mathbf{d}
+16m\mathbf{c}_1 -4ig[\mathbf{A},b^{\,0}],\nn\\
\{\mathbf{D}\times,\mathbf{c}_1\}+\frac{i}{4}[\mathbf{D},e]&=&
2\left(-\mathbf{p} e + ig[\mathbf{A}\times,\mathbf{c}_1]\right).\lb{c1} \eea
The solution for a corresponding abelian QED case is known \cite{bgr}, but
the derivation of a corresponding non-abelian analog is rather
time-consuming work. We are limited here with a very simple case that have
no analogy in QED. Assuming that $\mathbf{A}^a=\mathbf{A}, \forall a$
 ("colour democracy"), we have
$\mathbf{H}=0$ and the system \re{c2},\re{c1} is reduced to
\be
[\mathbf{A}, b^0 ] = 0, \qquad [\mathbf{A} , \mathbf{b} ] = 0,\qquad
\mathbf{p} a = m\mathbf{b}, \ee all other components are zero. This system
allows the solutions with zero colour currents but with non-zero colour
charges, for example
\be
b^0 \sim a,\qquad a=a_1\hat 1+\mathbf{a}_2 \mathbf{A},\qquad
\mathbf{b}=a\mathbf{p}/m, \ee where $a_1, \mathbf{a}_2$ are arbitrary constants.

\subsection{Chiral limit}
The additional simplification is possible in the chiral limit $m\to 0$. In
this case, the equations for the vector components of the Wigner function
are separated from others. We suppose also the flux-tube symmetry in the
electric field direction $\mathbf{n}=\mathbf{E}/E$, then
 \begin{eqnarray}\label{flux} \mathbf{b}&=&
b\, \mathbf{n} + b_{\trans}\mathbf{p}_{\trans},\nn\\ \mathbf{d}&=
&d \cdot (\mathbf{n}\times\mathbf{p}_{\trans}),
\end{eqnarray}
 because of $\mathbf{b}$ is polar and $\mathbf{d}$ axial
vectors. By the initial conditions of the type \re{quasivac}, it follows
from Eq.\re{hom} that $d^{\,0}=0$ and $b^0=0$ (neutral system). Then we
obtain
\bea\label{chir} \p_t \mathbf{b}^s +\frac{g}{4}E^a\p_p\,\mathbf{b}^a &=&
2\mathbf{p}\times\mathbf{d}^s \nn\\
 \p_t \mathbf{b}^a+gE^a\p_p\,\mathbf{b}^s &=& 2\mathbf{p}\times\mathbf{d}^a ,\nn\\
\p_t \mathbf{d}^s+\frac{g}{4}E^a\p_p\,\mathbf{d}^a &=& 2\mathbf{p}
\times\mathbf{b}^s,\nn\\
 \p_t \mathbf{d}^a+ gE^a\p_p,\mathbf{d}^s &=& 2\mathbf{p}\times
\mathbf{b}^a,\nn\\
 f^{abc}\mathbf{A}^b \mathbf{b}^c&=&0,\nn\\ f^{abc}\mathbf{A}^b \mathbf{d}^c&=&0.
\eea
The two last algebraic equations play the role of
constraints and have the particular solution ("color democracy")
\be\label{coldem} \mathbf{A}^a=\mathbf{A},\qquad \mathbf{b}^a=\mathbf{b},
\qquad \mathbf{d}^{\,a}=\mathbf{d}, \qquad a=1,2,3.
\end{equation}
These conditions correspond to the special case of Abelian dominance
approximation \cite{gyul,elze2} in relation to the gluon  field. If we
assume that the conditions \re{coldem} are carried out for all others
components WF also, the solution get the form \re{quasivac}
\begin{equation}\label{eta}
W^a(t,p)=\eta\, W^s(t,p), \qquad a=1,2,3,
\end{equation}
where $\eta$ is some parameter. We find two admissible values after
substituting that in the system \re{chir}
\begin{equation}\label{sollam}
  \eta= \pm \frac{2}{\sqrt{3}}.
\end{equation}
The system \re{chir} is reduced to the three scalar equations at the account
of the representation \re{flux}
\begin{eqnarray}\label{chirs}
\p_t b_{\longi} +\eta\,g E\p_p\, b_{\longi} &=& 2 p^2_{{\trans}} d ,\nn\\
\p_t b_{{\trans}} +\eta\, g E\p_p\, b_{{\trans}} &=& -2 p_{\longi} d ,\nn\\
\p_t d +\eta\,g E\p_p\,d &=& 2( p_{\longi}\, b_{{\trans}}-b_{\longi}).
\end{eqnarray}
This equation set can be solved numerically by the characteristics method.
The remaining part of Eq.\re{hom} is reduced to
  \begin{eqnarray}\label{rem}
\p_t a +\eta\, g E\p_p\, a &=& \mathbf{p}\,\mathbf{c}_1, \nn\\
\p_t \mathbf{c}_1 +\eta\, g E\p_p\, \mathbf{c}_1 &=& -\mathbf{p}\,a ,\nn\\
\p_t \mathbf{c}_2 +\eta\, g E\p_p\, \mathbf{c}_2 &=& -\mathbf{p}\,e, \nn\\
\p_t e +\eta\, g E\p_p\,e &=& 4 \mathbf{p}\,\mathbf{c}_2.
\end{eqnarray}
We find two integrals of motion combining these equations in pairs
  \begin{eqnarray}\label{int}
D_t( a^2+ 4\mathbf{c}_1^2)&=&0,\quad\to\quad  a^2+ 4\mathbf{c}_1^2=const,\nn \\
D_t( e^2+ 4\mathbf{c}_2^2)&=&0,\quad\to\quad  e^2+ 4\mathbf{c}_2^2=const,
 \end{eqnarray}
here $ D_t=\p_t +\eta\,g E\p_p\ $. It follows for the  field-free initial
conditions that $ e=0$ and $ \mathbf{c}_1=0$. The self-consistent evolution of
the mean gluon field obeys the Yang-Mills equation
\begin{equation}\label{ym}
  \dot{E}=-2g\int\, \frac{d^3p}{(2\pi)^3}\, b.
\end{equation}
The equations \re{chirs}, \re{rem} and \re{ym} are the closed system for the
numerical investigation of initial value problems such as
the pair creation in strong field with the account of a
back-reaction of the produced particles on the evolution of the mean
gluon field.

The simple solutions of the type \re{eta} do not satisfy to the Eq.\re{hom3}
for the SU(3) case. But the reduction to the ordinary differential equation
is possible, nevertheless, at use the more complicated representation of the
type \re{quasivac}
\begin{equation}\label{demsu3}
W^a(t,p)=\eta^a W(t,p), \qquad a=1,2\ldots N.
\end{equation}
We obtain the non-linear equations system for the admissible values of
$\eta^a$ by substituting these relations in Eqs.\re{hom3}
\begin{equation}\label{eta3}
\eta^a \left(\sum_i \eta^i\right)=6\,
\left[1+\sum_i d^{\,aik}\eta^k\right],
\qquad a=1,2\ldots 8,
\end{equation}
where $d^{\,abc}$  has three independent non-zero values only
\cite{ind}.

\section{Comparison with QED}\label{qed}

The equations \re{proj_a}-\re{proj_e} can be transformed to QED case by formally
setting the color matrices $t^a$ equal to the unit one. Then the
commutators of the gauge fields with the spinor components vanish, whereas
the anti-commutators give a factor 2:
\bea\label{qed_a}
\p_\tau a+gE_\alpha \p^\alpha_{\trans} (p)a &=&
4p^\alpha_{\trans} c^{\trans}_\alpha,\nn \\
% b_parallel
\lb{qed_b_par}
\p_\tau b^\parallel+\p^\alpha_{\trans} (x) b_\alpha^{\trans}
+g\p^\alpha_{\trans} (p)\left(b^\parallel E_\alpha
- b^\beta_{\trans} F_{\alpha\beta}^{\trans} \right)&=&0, \nn\\
% b_perp
\lb{qed_b_perp}
\p_\tau b^\mu_{\trans} - \p_{\trans}^\mu (x) b^\parallel +
+g\p^\alpha_{\trans} (p)\left(b^\mu_{\trans} E_\alpha +
b^\parallel F_{\alpha }^{{\trans} \mu}\right)&=&
2p_\alpha^{\trans} d_\beta^{\trans}\varepsilon^{\,\alpha\beta\mu}+4mc_{\trans}^\mu
 \nn\\
% tensor_parallel
\p_\tau c_{\trans}^\mu + \p_\alpha^{\trans} (x) c^{\alpha\mu}_{\trans}
+g\p^\alpha_{\trans} (p)\left(E_\alpha c_{\trans}^\mu -
F^{\trans}_{\alpha\beta} c_{\trans}^{\beta\mu}\right)&=&
p_{\trans}^\mu a- m b_{\trans}^\mu,\nn\\
%tensor_perp
\p_\tau c^{\mu\nu}_{\trans}+\p^\nu_{\trans} (x) c_{\trans}^\mu-
\p^\mu_{\trans} (x) c_{\trans}^\nu &+&\nn\\
+g\p^\alpha_{\trans}(p)\left(E_\alpha c^{\mu\nu}_{\trans} +
F_{\alpha}^{{\trans}\mu} c_{\trans}^\nu
- F_{\alpha}^{{\trans}\nu} c_{\trans}^\mu \right)&=&
e p_\alpha^{\trans} \,\varepsilon^{\alpha\mu\nu}, \nn\\
% axial parallel
\p_\tau d^\parallel +\p_{\trans}^\alpha (x) d^{\trans}_\alpha +
g\p^\alpha_{\trans}(p)
\left(d^\parallel E_\alpha - F_{\alpha\beta}^{\trans} d^\beta_{\trans}
\right) &=&2me, \nn\\
% axial perp
\p_\tau d^{\,\mu}_{\trans} - \p_{\trans}^\mu (x) d^{\,\parallel}+
g\p^\alpha_{\trans}(p)\left(d^{\,\mu}_{\trans} E_\alpha +
F_{\alpha}^{{\trans}\mu} d^{\,\parallel}\right)&=&
2p_\alpha^{\trans} b_\beta^{\trans}\varepsilon^{\,\alpha\beta\mu},\nn\\
% pseudo-scalar
\p_\tau e+gE_\alpha \p^\alpha_{\trans} (p)e  +2md^{\,\parallel}&=&
-2p_\alpha^{\trans} c_{\beta\delta}^{\trans}\varepsilon^{\,\alpha\beta\delta}
.\lb{qed_e}
\end{eqnarray}
These equations are corresponds the formulae (4.9) - (4.16) of the work
\cite{hmr} (part 2). The system \re{qed_e} can be reduced for the simple
field \re{electr} to three scalar ordinary differential equations
\cite{bgr}, which allows the simple numerical investigation.

\section{Summary}\label{summ}

We have derived the system of KE for description of quark-antiquark plasma
created from vacuum under action of a strong quasi-classical gluon field.
The single-time Wigner function formalism allows in contrast to other
approaches of this kind to formulate correctly the Cauchy problem. It is
particularly important for non-perturbative description of vacuum particle
creation.

We have analyzed some special cases of obtained system of KE (vacuum
solution, space-homogeneous time dependent color electric field e.t.c.). It
is shown that KE is complex system of the partial differential equations
even in the most simple case (chiral limit). That system is rather difficult
for the numerical investigation, while the KE in QED allows the reduction to
the set of ordinary differential equations \cite{bgr,grib} for some field
configurations. Thus the transition to QCD either sets higher request to
level of computer calculations or needs a some additional non-perturbative
model assumptions.

\subsection*{Acknowledgments}
We thank A. Reichel for helpful discussions. This work was supported partly
by the Ministry of Education of the Russian Federation under grant N
E02-3.3-210 and Russian Fund of Basic Research (RFBR) under grant
03-02-16877.

\end{document}